\documentclass[twocolumn,twoside,slac_two]{revtex4}
\usepackage{aas_macros}
\usepackage{graphicx}
\usepackage{wasysym}
\usepackage{fancyhdr}
\newcommand{\pics}{./}
\begin{document}

\title{Indirect Visibility of Gravitational Waves in Magnetohydrodynamic Plasmas}

\author{J. Moortgat}
\affiliation{Department of Astrophysics, University of Nijmegen, The Netherlands}
\affiliation{Theoretical Astrophysics, California Institute of Technology, Pasadena, USA}
\author{J. Kuijpers}
\affiliation{Department of Astrophysics, University of Nijmegen, The Netherlands}

\begin{abstract}
We propose a mechanism to make gravitational waves (GWs) visible in the electromagnetic domain.
Gravitational waves that propagate through a strongly magnetized plasma interact with the plasma through its anisotropic stress-energy tensor and excite magnetohydrodynamic (MHD) wave modes. 
In catastrophic events such as the merger of a double neutron star  binary, a large fraction of the total binding energy of the system is released in the form of GWs observable by LIGO, and the amount of
energy transferred to the MHD waves is substantial. 
These modes, however, are excited at the same frequency as the GW and are not directly observable. In this paper we investigate radiation processes that operate in the presence of the gravitationally excited MHD waves and radiate in the radio regime accessible to LOFAR. We present order of magnitude estimates for the spectral flux of a merger detectable by a LOFAR.
\end{abstract}

\maketitle

\thispagestyle{fancy}

\section{INTRODUCTION}
Gravity is the weakest of all the fundamental forces and, not surprisingly, gravitational waves have proven to be the most elusive signals in astronomy. In fact, now that almost every  electromagnetic frequency is observable and even cosmic ray and neutrino astronomy are becoming a reality, the first direct detection of GWs still has to be made. With the development of advanced resonant bar and in particular interferometric GW detectors, this will most likely happen within the next one or two decades. The same property of GWs that makes them so hard to detect is their main advantage as a diagnostic of the Universe: because GWs hardly interact with anything the entire Universe is transparent to them.
The planned GW detector in space, LISA \cite{lisa} expects to detect the signal from merging supermassive black holes out to any distance and LIGO \cite{ligo} which is presently under construction
 could probe for instance the central engine of gamma-ray bursts (GRBs) which is opaque to electromagnetic radiation. 

In general, electromagnetic and GW observations are very complimentary and an electromagnetic counterpart of any GW signal and vice versa is invaluable. As an example: with electromagnetic observations it is very difficult to determine the inclination of a binary system, whereas this could be easily measured from the polarization of a GW signal \cite{schutz96}.
On the other hand, the amplitude of a GW measured on earth does not allow a unique determination of the distance and the intrinsic strain at the source (which can be measured only from the chirp of the system). When the distance can be determined electromagnetically, this uncertainty can readily be removed.

Rather than studying different types of GW sources and trying to find  accompanying electromagnetic processes, we investigate whether the GW themselves produce an electromagnetic signal when they propagate through the magnetohydrodynamic (MHD) plasma in which many GW sources are embedded. In particular one can think of 
 rapidly spinning NS with small ($<10^{-5}$) asymmetry,
low-mass X-ray binaries,
neutron stars with an $r$-mode instability,
 asymmetric supernova core collapse and bounce (GRB with afterglow),
newly born neutron stars that boil and oscillate (magnetars) and
 coalescing compact binaries (short GRB candidates).
Gravitational waves only interact to linear order with mediums that have an anisotropic stress-energy tensor. Interaction is therefore possible with a magnetized MHD plasma and not, for instance, with dust or an isotropic hydrodynamic ideal fluid.

We have found in previous work (see \cite{moortgatI}, \cite{moortgatII} and references therein) that $\times$ and $+$ polarized GW couple to Alfv{\'e}n and fast magneto-acoustic MHD modes, respectively. Here the $\times$ polarization refers to GWs that are polarized at an angle to a uniform magnetic field. In the limit of a strongly magnetized tenuous plasma these MHD waves can interact coherently with a GW and grow linearly with distance up to a maximum value.
 The frequency of the excited MHD modes is, however, the same as that of the GW and therefore lower than the interstellar plasma frequency and not observable directly. 

In this paper we investigate subsequent inverse Compton (IC) scattering of the low frequency (kHz) MHD wave on the relativistic electrons (and positrons) in the wind, either the bulk secondary particles or the even more relativistic primary particles. Inverse Compton scattering on electrons and positrons with a Lorentz factor of $\gamma > 100$ boosts the frequency of the radiation to the regime of the LOFAR \cite{lofar} low frequency radio array currently under construction in the Netherlands.
Finally, we mention the possibility of coherent radiation when the MHD mode acts as an undulator for the relativistic electrons similar to the free electron laser process \cite{fung}.

The set-up of this paper is as follows: Section~\ref{sec::sec1} summarizes some of the main conclusions of \cite{moortgatI} and \cite{moortgatII}, Section~\ref{sec::sec2} elaborates on the specific situation of GWs propagating through the magnetosphere and wind of a double neutron star (neutron star) binary close to merging. In Sections~\ref{sec::sec3} -- Section~\ref{sec::sec4} we derive a rough estimate for the spectral flux on earth as a result of the inverse Compton scattering, taking into account possible damping mechanisms and propagation effects and we discuss some issues that are subject of future research such as the coherent radiation mechanism. We end with conclusions in Section~\ref{sec::sec5}.

\section{\label{sec::sec1}MHD MODES EXCITED BY GW}
The only oscillating components $\delta T^{ab}$ of the stress-energy tensor in the rest frame of a perfect magnetofluid that couple to the GW and can not be removed by a gauge transformation depend on the magnetic field.
Explicitly, the Einstein field equations reduce to evolution equations for the GW:
\begin{eqnarray}\label{eq::gwevolutionx}
\Box h_+  (z,t) &=&  4 B_x^0 \delta B_x (z,t) , \\\label{eq::gwevolutiony}
 \Box h_\times (z,t)  &=& 4 B_x^0 \delta B_y (z,t),
\end{eqnarray}
where the ambient magnetic field is chosen to lie in the $x$-$z$ plane: $\vec{B} = \vec{B}^0 + \delta \vec{B}$ and $\vec{B}^0 = B^0 (\sin\theta, 0, \cos\theta)$.

Approximating the GW as a driver, the evolution equations for the magnetic field can be solved and look like:
\begin{equation}\label{eq::bx}
\delta B_x \propto \frac{1}{2} h_+ B_x^0, 
\qquad 
\delta B_y \propto \frac{1}{2} h_\times B_x^0.
\end{equation}
These results are reminiscent of the spatial deviations of test masses in interferometers such as LIGO ($\delta x = \frac{1}{2} (h_+ x_0 + h_\times y_0)$ and
$\delta y = \frac{1}{2} (h_\times x_0 - h_+ y_0)$).

\begin{figure}[!h]
\begin{center}
\includegraphics[width=65mm]{\pics/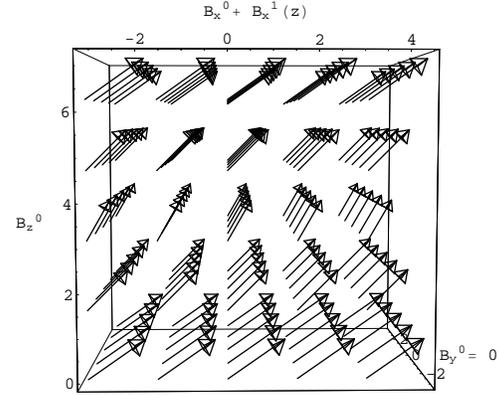}
\caption{Magnetic field in the magneto-acousic mode. The perturbation is exaggerated to emphasize the overall behavior (arbitrary units).}
\label{fig::mswfield}
\end{center}
\end{figure}

The solution for $\delta B_x$ is illustrated in Figure~\ref{fig::mswfield} and corresponds to a compressional fast magnetosonic wave (MSW) with both electromagnetic and gas properties. Coherent interaction with the GW is possible when the phase velocity of the MSW approaches that of the GW. In a Poynting flux dominated plasma where the Alfv\'en velocity $u_\mathrm{A}$ is relativistic and much larger than the sound velocity, this limit is satisfied and the perturbations are allowed to grow linearly with distance:
\begin{eqnarray}
\delta B_x (z,t) &\simeq& \frac{h_+}{4\gamma^2} B^0 \sin\theta\ \omega z\  \Im\left[\mathrm{e}^{i\omega (z-t)} \right].
\end{eqnarray}
The phase velocity of the slow MSW is always much smaller than the fast mode, so it can never interact coherently with the GW.

\begin{figure}[!h]
\begin{center}
\includegraphics[width=65mm]{\pics/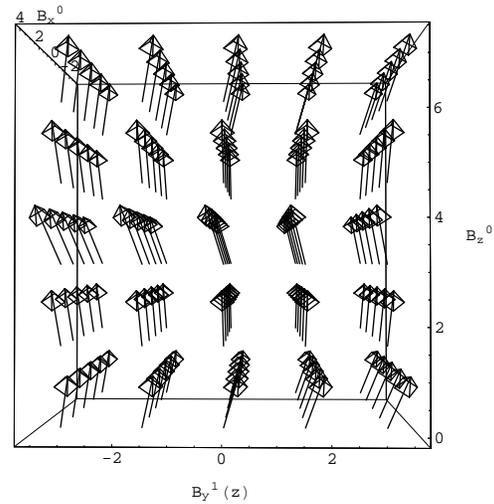}
\caption{Magnetic field in the Alfv{\'e}n mode.}
\label{fig::alfvenfield}
\end{center}
\end{figure}

The second expression in Eq.~\ref{eq::bx} corresponds to non-compressional shear Alfv\'en waves, illustrated in Figure~\ref{fig::alfvenfield}. The condition for coherent interaction with the GW is more stringent because its phase velocity $u_{\mathrm{A}\|} = u_\mathrm{A} \cos\theta$ has to approach the velocity of light, but at the same time its amplitude is $\propto B^0_x \propto u_\mathrm{A} \sin\theta$. Therefore, in the case of coherent interaction the amplitude of the Alfv\'en waves is suppressed by a small factor $\theta \ll 1$.

\begin{equation}\label{eq::Alfvengrowth}
\delta B_y (z,t) \simeq \frac{h_\times}{4\gamma^2} B^0 \theta \  \omega z\ \Im[\mathrm{e}^{i \omega (z-t)}] + {\mathcal O}[\theta^2].
\end{equation}

\begin{figure}[!h]
\begin{center}
\includegraphics[width=60mm]{\pics/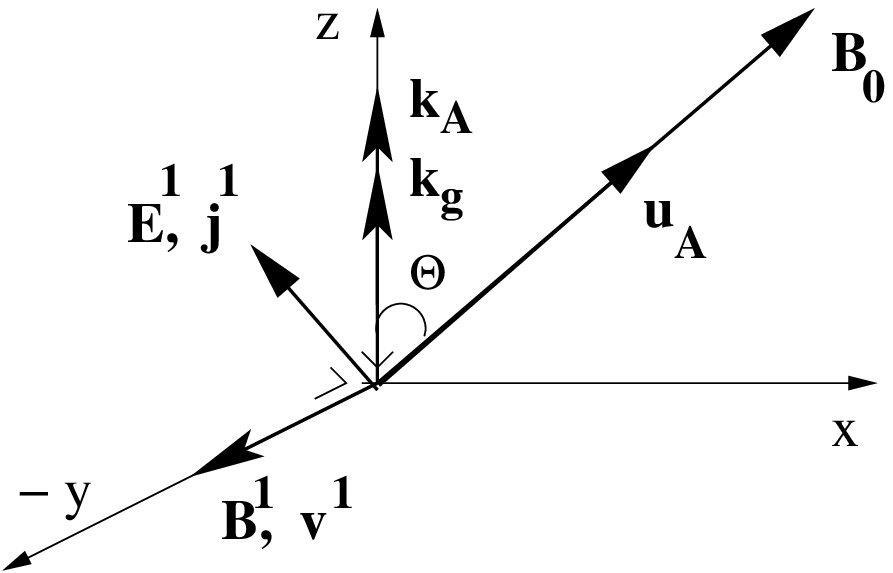}
\includegraphics[width=60mm]{\pics/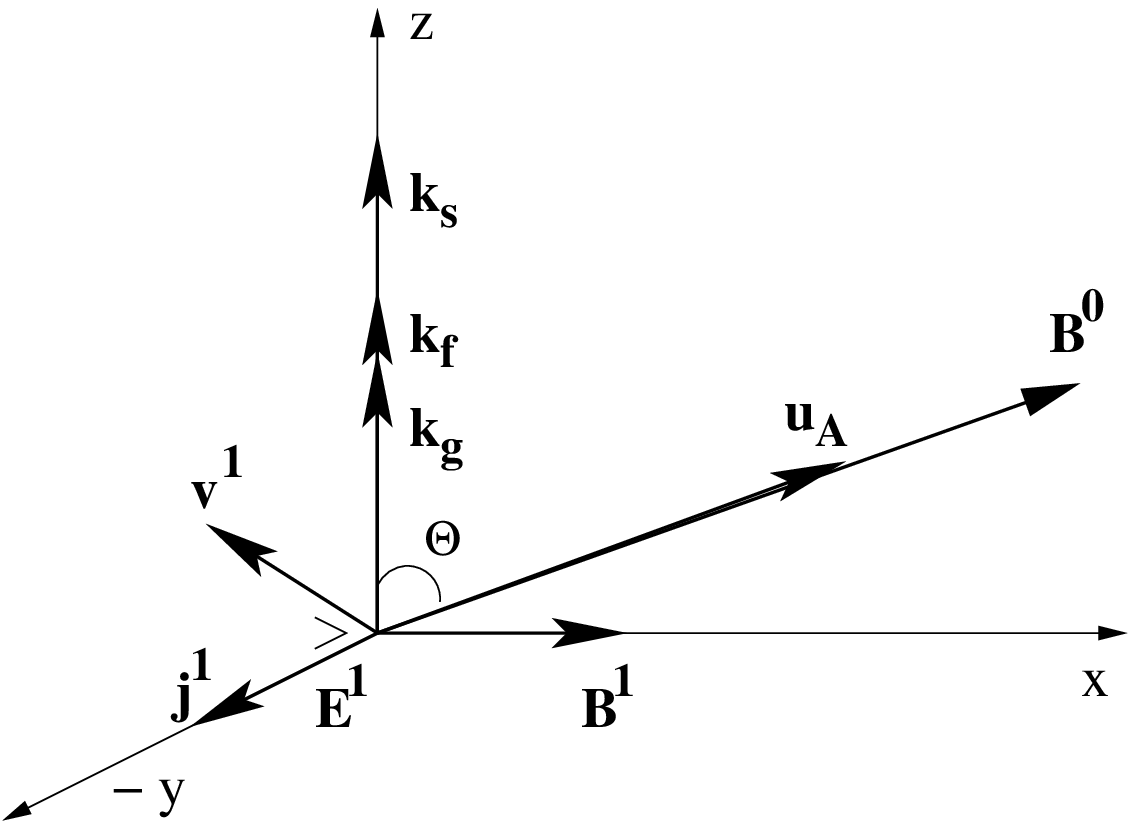}
\caption{Polarization of Alfv{\'e}n (top) and fast magnetosonic modes (bottom).}
\label{fig::polarizations}
\end{center}
\end{figure}
In fact the polarization of the GWs emitted by a binary merger 
is linear only for an edge-on and circular for a face-on binary with the same sense as the binary motion on the sky. If the GW is for instance right circularly polarized, 
simply replace $h_{+}\rightarrow h_{\mathrm{R}}/\sqrt{2}$.

As was mentioned in the previous section, the GW only interacts directly with the magnetic field. In particular the plasma motion in a GW is generally non-compressional. However, in a perfectly conducting plasma the particles are strongly coupled to the magnetic field lines and Maxwell's equations couple to the matter conservation laws through the current density. 
Consequently, perturbations in pressure, density, magnetic field gradients, currents and a drift velocity are also excited in the MSW, whereas the non-compressional Alfv\'en waves cause a divergence of the electric field and a corresponding charge density fluctuation. The orientation of these different components of the MHD modes, i.e. the polarizations, are shown in Figure~\ref{fig::polarizations}.

\section{\label{sec::sec2}GENERAL PICTURE}
\begin{figure}[h!]
\begin{center}
\includegraphics[width=75mm]{\pics/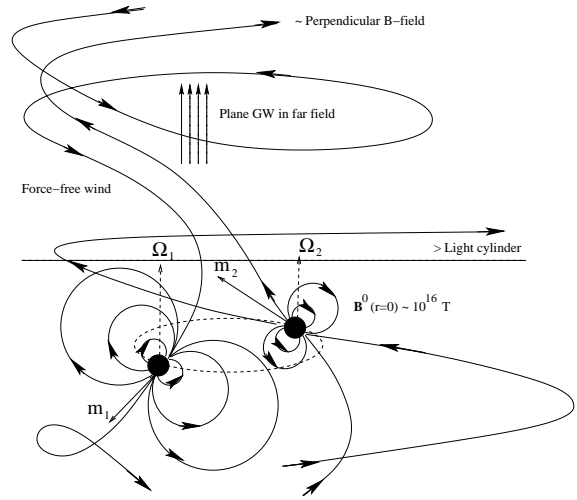}
\caption{Magnetic field configuration of a double neutron star binary merger.}
\label{fig::fig1}
\end{center}
\end{figure}

We consider a merger of two neutron stars where each neutron star has its own magnetic field that falls off as a dipole close to the surface: $ B (r) = B_\star \left(R_\star/r\right)^3$.
As the merger coalesces the orbital frequency increases and dominates over any other (rotational) motion.
Therefore, we assume that at the end of the spiral-in phase the \emph{orbital} rotation of the binary (with $\Omega_\mathrm{b} \sim 10^3$ rad/s) determines the \emph{light-cylinder} radius: $R_\mathrm{lc} = c/\Omega_\mathrm{b} \simeq 320$ km. At the light cylinder plasma would require superluminal velocities to corotate.  
A steady charge density cannot be maintained at the \emph{Goldreich-Julian density}, $n_\mathrm{GJ} = \vec{2 \epsilon_{0}\Omega}_\mathrm{b}\!\cdot\!\vec{B}/e$, everywhere. 
As a result, a strong electric field develops along the open magnetic field lines above the {\em polar cap} and charged `primary' particles are extracted from the surface with a density $n_\mathrm{p} \simeq n_\mathrm{GJ\star}$ and accelerated to high Lorentz factors (a typical number is $\gamma_\mathrm{p} \sim 10^7$). The available potential jump is proportional to $\Delta \Psi \propto \Omega_\mathrm{b} B_\star $ in our case and can be much larger than for single pulsars since the orbital frequency becomes large for neutron stars of arbitrary field strengths.

Processes such as curvature radiation and inverse Compton emission then result in a cascade of  `secondary' $e^\pm$ pairs with a particle number density $n_\mathrm{s} = 2 M n_\mathrm{p}$, where $M$ is called the \emph{multiplicity}. Due to energy conservation $n_\mathrm{p}\gamma_\mathrm{p} = n_\mathrm{s}\gamma_\mathrm{s}$, so the Lorentz factor of the secondary particles is $\gamma_\mathrm{s} = \gamma_\mathrm{p}/(2 M) \sim 100$ for $M \sim 10^5$. The secondary plasma, probably interspersed with beams of primary particles, flows out as a relativistic wind along the open magnetic field lines which develop into spirals further out since the toroidal component of the field soon dominates the poloidal component. The wind remains \emph{force-free} up to a large distance.

As the binary looses angular momentum through gravitational radiation it spirals in and eventually the two neutron stars merge and form a black hole. In the process, a large fraction of the binding energy of the system is released in the form of GWs that propagate through the surrounding plasma and excite MHD waves. In the relativistic wind the dominant mode will be the fast magnetosonic wave that propagates with a highly relativistic Alfv{\'e}n phase velocity $u_\mathrm{A} \uparrow c$, overtaking the particles further out (that are themselves relativistic). 

In the next section we will discuss how the energy transferred to the plamsa by the GW can be released in the form of inverse Compton radiation.

\section{\label{sec::sec3}INVERSE COMPTON RADIATION}
The relativistic flow in the jet cannot remain force free after the GW
pulse has travelled through the jet and excited the MHD waves which -- by
definition -- impose forces and accelerations everywhere on the local
plasma. As a result the jet is expected to radiate. For simplicity we
assume that the particles are characterized by an outflow of secondary
particles at multiplicity $M$ and a random motion with Lorentz factor
$\gamma_s$ superimposed on a systematic outflow with a similar Lorentz
factor. Of course, there will also be the remnant of the primary beam with
Lorentz factor $\gamma_p$ but these particles will not radiate at radio
frequencies. We note however that the secondary particles will (see 
Eq.~\ref{eq::resonance} below). Actually, the radiation from the secondary particles
will be partly synchrotron in the magnetic field of the MHD waves --
largely at frequencies above the LOFAR band -- and partly inverse Compton
radiation. Although the inverse Compton process is fixed by the frequency
of the GW and the Lorentz factor of the particles -- and has therefore the
same frequency and in the radio domain all over the jet -- this radiation
can only escape sufficiently far out where the relativistic plasma
frequency has fallen below the frequency of emission $\omega_s > \omega_p
/\sqrt{ <\gamma ^{3}>} \approx \omega_p /\sqrt{ \gamma_s}$ where $\omega_p$ is the
non-relativistic plasma frequency of secondary particles in the observer
frame and the average is over the energy distribution of the secondary
particles \cite{melrose99}. 

\begin{figure}[h!]
\begin{center}
\includegraphics[width=40mm]{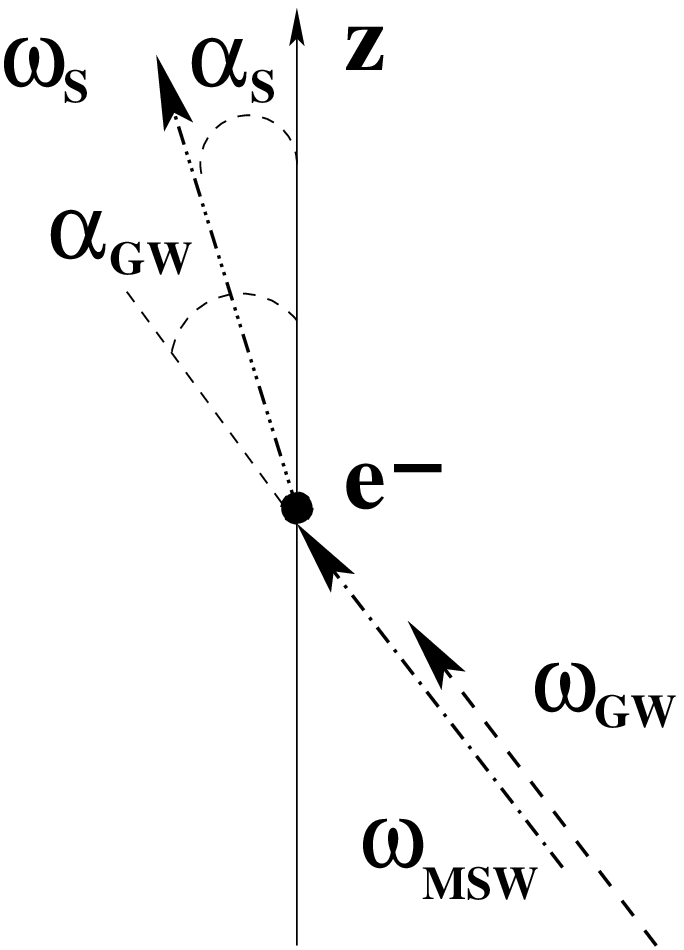}
\includegraphics[width=40mm]{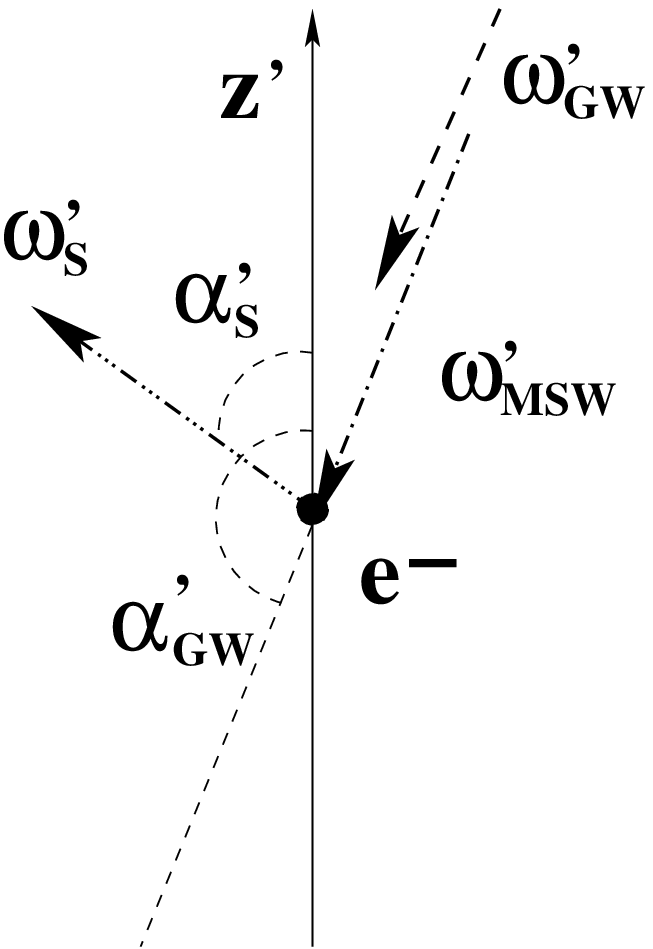}
\caption{Inverse Compton process in plasma (left) and lab (right) frame.}
\label{fig::iclabplasma}
\end{center}
\end{figure}

Inverse Compton scattering is a natural radiation process because the GW -- and therefore also the MHD wave -- propagate outwards radially, whereas the wind is collimated by the background magnetic field and has an essentially cylindrical geometry. Consequently, the MHD wave overtakes the wind at an angle, and in fact the relativistic electrons will see it approaching almost head-on in their rest frame (Figure~\ref{fig::iclabplasma}). In fact, since we are studying the regime of a strongly magnetized tenuous plasma, the MHD wave behaves similar to a vacuum electromagnetic wave propagating at the speed of light. We will therefore treat the MHD wave simply as incident photons with $\omega = \omega_\mathrm{mhd} = 
\omega_\mathrm{gw}$ in a first estimate. 

\begin{figure}[h!]
\begin{center}
\includegraphics[width=75mm]{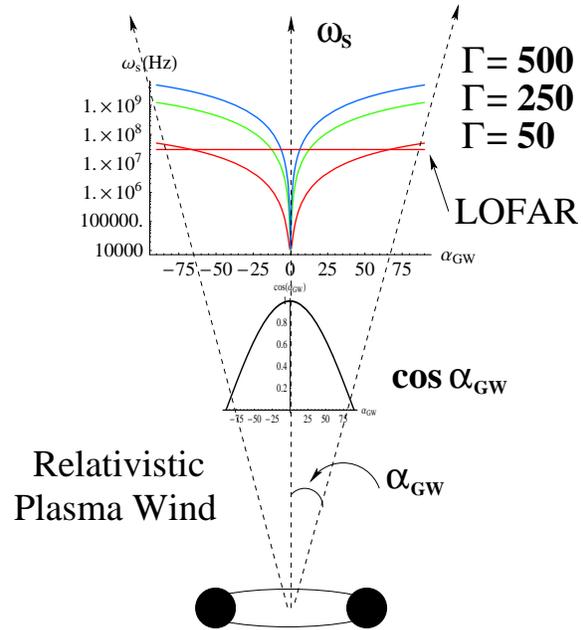}
\caption{Scattering of the radial MHD waves on the cylindrical flow in the wind (in this illustration the angle $\alpha_\mathrm{GW}$ is compressed).}
\label{fig::collimated}
\end{center}
\end{figure}

The frequency of the scattered radiation can most readily be derived from the resonance condition:
\begin{equation}\label{eq::resonance}
\omega_\mathrm{mhd} - k_\mathrm{mhd} \beta \cos\alpha_{\mathrm{gw}}= \omega_\mathrm{s} - k_\mathrm{s} \beta \cos\alpha_\mathrm{s}  \simeq \frac{\omega_{\mathrm{s}}}{2\gamma_\mathrm{s}^2}
\end{equation}
where the last equality is for the photons scattered in the forward direction ($\alpha_\mathrm{s} =0$). The frequency of the scattered radiation as a function of the angle between the incident MHD wave and the bulk velocity in the wind is plotted in Figure~\ref{fig::collimated}. Already for relatively small angles $\alpha_{\mathrm{gw}}$ does the boosted frequency fall into the LOFAR regime, which is $10$ -- $220$ MHz.

The power density in inverse Compton radiation is given by \cite{ryblight}
\begin{equation}\label{comptonpower}
P_\mathrm{ic} \sim \kappa \gamma_\mathrm{s}^2 \beta^2 \sigma_\mathrm{T} \epsilon_\mathrm{EM} n_e,
\end{equation}
where $n_e$ is the density of scattering electrons (in fact, the scattered radiation from the electrons and the positrons adds constructively). $\beta$ is the velocity of the electrons and $\gamma_\mathrm{s}$ the corresponding relativistic Lorentz factor, $\epsilon_\mathrm{EM}$ is the energy density in the incident MHD waves and $\kappa$ is a geometrical factor of order unity, which depends on the angular distribution of the incident radiation ($\kappa = 4/3$ for an isotropic radiation field). Note that the Thomson cross-section, $\sigma_{\mathrm{T}}$, is only a rough estimate for the scattering of a MHD wave on relativistic charged particles.

The energy density in the MHD mode (with $\delta E^2 + \delta B^2 \simeq 2 \delta B^2 \gg \frac{1}{2} \rho \delta v^{2}$) is \cite{moortgatII}:
\begin{eqnarray}
\epsilon_\mathrm{EM} &=& \frac{|\delta \vec{B}|^2}{\mu_0} = \left[\frac{(B_x^0)^2}{2 \mu_0}\right] \frac{(k_\mathrm{mhd} z)^2 h_{+}^2 }{2 \gamma^4}.
\end{eqnarray}
For a GW burst of duration $\Delta t\leq 0.1$ s, $(k_\mathrm{mhd} z_\mathrm{max})^2 h_{+}^2 = \omega^2 h_{+}^2 \Delta t^2 = \mathcal{O}[1]$ 
and the excited MHD wave amplitudes can be quite significant for a mildly relativistic wind. Note also that in the analytical calculations we have assumed a uniform constant background magnetic field. In the numerical estimates in the next section, we will take account of the $1/r$ dependencies in a WKB approach so that  the MHD wave amplitude falls of as $1/r$ within the interaction region just like the background field.

\subsection{Numerical estimates}

Radiation from the entire jet will be emitted as soon as the GW pulse has
passed through and the MHD waves are excited. The burst of electromagnetic
radiation will be compressed to a duration $\Delta t \approx L/ (c 2
\gamma_s^2)$ as is familiar from GRB models. However, even though the
entire jet is radiating at the same inverse Compton frequency the jet
expands and is therefore very inhomogeneous. In our estimate of the total
power in inverse Compton radiation we will just approximate the jet as
being built up out of different parts each of uniform properties and
extent equal to the local scale height $L \approx r$. 

In the absence of
collimation both the electron density, the backgound magnetic energy
density and the MHD wave field energy density fall off as $1/r^2$ and
the
simple estimate of Eq.~\ref{comptonpower} integrated over $r^{2}dr$ shows that the power falls of with $1/r$.
Radiation in the LOFAR radio domain can only escape far away from the
light cylinder where the plasma density has decreased sufficiently. 
 Note
however that a mild collimation of the jet with $ B \propto 1/\sqrt{r}$ so
that both $\epsilon_{EM}, n_e \propto 1/r$ already leads to the opposite
result that the power increases ($\propto r$). 

\begin{table}[!h]
\begin{center}
$$
\begin{array}{|rcl|}
\hline \textbf{Parameter} && \textbf{Value} \\
\hline 
B_\star &\sim& 10^{12}\ \mathrm{T} \\
\Omega_\mathrm{b} &\sim& 10^{3} \ \mathrm{rad~s}^{-1} \\
\gamma_\mathrm{p} & \sim&10^7 \\
\gamma_\mathrm{s} & \sim&10^2 \\
 M &\sim& 10^5\\
 R_\mathrm{lc} = \frac{c}{\Omega_\mathrm{b}} &\sim& 3 \cdot 10^5 \ \mathrm{m}\\
 n_\mathrm{lc} = \frac{2 M \epsilon_{0}\Omega_\mathrm{b} B_\mathrm{lc}}{e} &\sim& 3 \cdot 10^{23}\ \mathrm{m}^{-3} \\
 h_\mathrm{in} &\sim&10^{-3} \\
 \sigma_\mathrm{T} &\sim& 66.5 \cdot 10^{-30}\ \mathrm{m}^{-2} 
\\\hline
\end{array}
$$
\caption{Numerical values in Eq.~\ref{totalcomptonpower}.}
\label{table::values}
\end{center}
\end{table}

When the jet is collimated and using the values in Table~\ref{table::values} we find:

\begin{eqnarray}\label{totalcomptonpower}\nonumber 
P^\mathrm{tot}_\mathrm{ic} &\sim& \kappa \gamma_\mathrm{s}^2 \beta^2 \sigma_\mathrm{T} 
\int^{2 r}_{r}\! 
\frac{\epsilon_\mathrm{EM} (R_\mathrm{lc}) n_e (R_\mathrm{lc}) R_\mathrm{lc}^2}{r^2}\ r^2 d r  \\
&\simeq& 4 \cdot 10^{30} \mathrm{W} \left(\frac{r}{R_{lc}} \right) ,
\end{eqnarray}
Coherent interaction of the GW with the MHD is only possible as long as their wave number mismatch is small, $(\Delta k) z \ll 1$, from which we calculated a maximum interaction length scale of $R_{\mathrm{max}} \sim 10^{15}\ \mathrm{m}$ in \cite{moortgatI}. Most of the inverse Compton radiation will be emitted in this last $L = R_{\mathrm{max}}$ with $P^\mathrm{tot}_\mathrm{ic} \sim 10^{40}$ W, which is approximately $10^{-5}$ of the power in GW!

For a binary merger close enough for a LIGO detection, i.e. in the Virgo cluster at $D\sim 1$ Gpc, this would yield a flux on earth of $2 \cdot 10^6\ \mathrm{Jy}$ at $30$ MHz assuming that the signal in reality is not monochromatic, but is incoherent with a bandwidth of $\Delta \nu \sim \nu = 30$ MHz. The duration of this transient signal is relatively short: $\Delta t = L/(2 \gamma_s^2 c) \approx 3$ min, 
but much longer than the cosmic ray events that LOFAR will also detect.

\section{\label{sec::sec4}DISCUSSION}
\subsection{Propagation effects}
Whether this signal is sufficiently strong, depends on a number of complicating factors. Firstly, the galaxy is quite bright at these low radio frequencies. From \cite{cane79} we find that at $30$ MHz the {\em galactic background} brightness temperature is $1.7 \cdot 10^4$ K. This is, however, still below LOFAR's antenna temperature of $\sim 2.3  \cdot 10^4$ K at the same frequency and well below our signal (both are approximately all sky temperatures and can be therefore be compared directly \cite{pacholczyk}). The galactic background temperature drops with frequency and it is even less of a problem at higher frequencies.

Secondly, {\em dispersion} is important at low frequencies. Since the radio signal is propagating through both interstellar and intergalactic plasma rather that a vacuum, its dispersion relation becomes frequency dependent. Radiation at lower frequencies is delayed with respect to higher frequencies and a single pulse is `smeared out' over a longer duration and with a correspondingly lower signal strength. 
This differential delay depends on the electron density, $n_\mathrm{e}$, integrated along the line of sight to a distance $D$, e.g. the electron column density or {\em dispersion measure} {\sc dm} in pc m$\!^{-3}$, so that the frequencies in Eq.~\ref{dm} are in MHz (and consequently also the differential dispersion is in s MHz$\!^{-1}$).
\begin{eqnarray}
\frac{dt}{d\nu}  &\sim& - 8\cdot10^{-3} \frac{\mbox{\sc dm}}{\nu^3},\\\label{dm}
\mbox{\sc dm} &=& \int_0^D n_\mathrm{e} dl,
\end{eqnarray}
When looking at a neutron star binary merger the signal is dispersed both by the host galaxy and by the intergalactic medium. Since these binaries close to a merger are very old systems that have received a randomly oriented kick during the birth of each neutron star, we expect them to lie relatively high above the galactic plane where the galactic dispersion is low. The dominant dispersion then occurs during intergalactic propagation and results in a dispersion measure of {\sc dm}$\sim 10^{7}$ pc m$\!^{-3}$ for a source at a distance of 
$100$ Mpc and an upper limit for the unknown intergalactic electron density of $n_\mathrm{e} \sim 0.1\ \mathrm{m}^{-3}$. In terms of delays this means
\begin{equation}
\begin{array}{lclcl}
\frac{dt}{d\nu} &\sim& - 3\ \mathrm{s}\ \mathrm{MHz}^{-1} \quad &\mbox{@ }& \nu = 30\ \mathrm{MHz},\\
&\sim& - 0.08\ \mathrm{s}\ \mathrm{MHz}^{-1} \quad &\mbox{@ }& \nu = 100\ \mathrm{MHz}.
\end{array}
\end{equation}
For comparison: the dispersion measure in a typical galactic medium with $n_\mathrm{e} \sim 3\cdot 10^{4}\  \mathrm{m}^{-3}$ over a distance of $D\sim 3.5\cdot 10^3 $ pc is {\sc dm} $= 10^{8}$ pc m$\!^{-3}$, so if the merger occurs deeper inside the host galaxy, or the line of sight crosses our own galaxy close to the disc the dispersion is an order of magnitude larger.

The effect of the dispersion on the signal will be to smear out the pulse 
within the observing band of $4$ MHz over say a minute. As a result the signal to noise will go down with the inverse square root of the increase in duration, but in view of the above estimates well within the 
sensitivity of LOFAR.

\subsection{Coherent radiation}

Here we mention the posibility of coherent radio emission from the jet.  
Once the fast mode MHD waves travel away from the jet axis into weaker
magnetic fields their amplitude relative to the ambient magnetic field
increases and may lead to the effective formation of a large amplitude
wiggler or undulator. Just as in the Free Electron Laser the fast particle
beam of secondary particles emits coherent radiation, again at the
frequency determined by Eq.~\ref{eq::resonance} but with an increased intensity (see \cite{fung}).

\section{\label{sec::sec5}CONCLUSION}
 We have shown that the pulse of GWs which is excited
in a merger sets up a pulse of fast mode outgoing MHD waves of the same
frequency in the pre-existing jet structure of the binary. The interaction
between fast (secondary) particles and these MHD waves creates radio
emission by the inverse Compton process which is expected to be observable
with LOFAR. The radio emission pulse comes from the outer regions of the
jet and is expected to be smeared out to longer duration by intergalactic
dispersion but still detectable with LOFAR. The inverse Compton signal can
serve as an indirect detection of GWs and be distinguished from other,
more conventional radio emission from the merger, by its low frequency
imprint. The best strategy to look out for these signals will be in GRBs
and future observations of GW detectors such as LIGO. The coupling between
GWs and MHD waves therefore offers in principle an important new and
indirect complementary detection technique of GWs.

\bigskip 
\begin{acknowledgments}
J. M. would like to thank the TAPIR institute at Caltech in general and Sterl Phinney in particular for their hospitality during his stay at Caltech.

This work was supported by NASA grant NNG04GK98G and the Leids Kerkhoven-Bosscha Fonds. 
\end{acknowledgments}

\bigskip 
\bibliography{moortgat_Stanford_poster}


%
\end{document}